\documentclass[twoside]{article}
\usepackage{zgwl}\usepackage{bm}
\usepackage{graphicx}

\newcommand\yj[1]{{$^*$}\parbox[t]{168mm}{#1}}
\newcommand\gj[1]{\parbox[t]{133mm}{\normalsize #1}}

\newcommand\bea{\begin{eqnarray}}
\newcommand\eea{\end{eqnarray}}

\catcode`@=11
\def\@evenhead{\vbox{\hbox to
\textwidth{\footnotesize\rm\thepage\qquad\qquad\ \ \hfill {1作 \sl
et al} \hfill Vol.\,17} \vs{1mm} \hbox to\textwidth{\hfill\vrule
depth0pt height0.15mm width\textwidth\hfill}}}
\def\@oddhead{\vbox{\hbox to \textwidth{\footnotesize\rm No.\,6 \hfill
Title ... \hfill\thepage} \vs{1mm} \hbox to \textwidth{\vrule
depth0pt height0.15mm width\textwidth}}}
\begin{document}
\setcounter{page}{1}
\thispagestyle{empty} \vs*{-1.2cm} \hbox
to\textwidth{\vbox{\footnotesize\bf\baselineskip=11pt\no Vol\ XX\
No\ XX,\ XXX\ XXXX\hfill\hs{1mm}{\normalsize\bf }
$\copyright$ XXXX\,\ Chin.\,\ Phys.\,\ Soc.\\
1674-1056/XXXX/XX(XX)/\pageref{first-page}-页数 \hfill\hskip
0.1mm{\normalsize\bf }
 {\huge\sf \hs{0.01mm}\makebox(1,9){Chinese Physics B\quad}}\hfill
and IOP Publishing Ltd}}
\par\noindent
\rule[3mm]{\textwidth}{0.2pt}\hs*{-\textwidth}\noindent
\rule[2.5mm]{\textwidth}{0.2pt}\label{first-page}
\vs{-.8cm} \bc {\LARGE\bf Reconstructing Dark Energy Potentials From
Parameterized Deceleration Parameters$^{*}$}
\footnotetext{\hs*{-.45cm}\fo\yj{Project supported by the National
Natural Science Foundation of China (Grant Nos 10573003 and
10703001), and Specialized Research Fund for the Doctoral Program of
Higher Education (Grant No 20070141034).}
\\\hs*{.8mm}$^\dag$E-mail:lxxu@dlut.edu.cn}
\footnotetext{\hs*{-4mm}\fo\sl  http://www.iop.org/journals/cpb　
http://cpb.iphy.ac.cn \hfill}
\vs{5mm}

{\rm Wang Yu-Ting,\ \ Xu Li-Xin$^\dag$,{\ \ L$\ddot{u}$ Jian-Bo,}\ \
and\ \  Gui Yuan-Xing
\\\vs*{2mm}

{\fo \sl
School of Physics and Optoelectronic Technology,\\
Dalian University of Technology, Dalian, Liaoning 116024, P. R.
China}

\vs{4mm}
\parbox[c]{155mm}{\parindent 20pt\fo
In this paper, the properties of dark energy are investigated
according to the parameterized deceleration parameter $q(z)$, which
is used to describe the extent of the accelerating expansion of the
universe. The potential of dark energy $V(\phi)$ and the
cosmological parameters, such as the dimensionless energy density
$\Omega_{\phi}$, $\Omega_{m}$, and the state parameter $w_\phi$, are
connected to it. Concretely, by giving two kinds of parameterized
deceleration parameters $q(z)=a+\frac{bz}{1+z}$ and
$q(z)=\frac{1}{2}+\frac{az+b}{(1+z)^2}$, the evolution of these
parameters and the reconstructed potentials $V(\phi)$ are plotted
and analyzed. It's found that the potentials run away with the
evolution of universe.

\vs{3mm}

\no{\normalsize \bf Keywords:}\ \gj{dark energy(DE), deceleration
parameter, reconstructed potential}

\no{\normalsize \bf PACC:}\normalsize\ 9880}} \ec \vs{3mm} \bmu

\section{Introduction}
Since 1998, the type Ia supernova (SNe Ia) observations [1] show
that the expansion of our universe is speeding up rather than
slowing down. During these years from that time, many additional
observational results, including current Cosmic Microwave Background
(CMB) anisotropy  measurement from WMAP [2], and the data of the
Large Scale Structure (LSS) [3], also strongly support this
suggestion. These observational results have received considerable
recognition. In order to understand the mechanism of the
accelerating expansion of the universe, the authors have done plenty
of work, which can be divided into two categories: on the one hand,
we can carry on the revision to the geometry items at the
left-handed side of the Einstein field
equation($R_{\mu\nu}-\frac{1}{2}g_{\mu\nu}R=4{\pi}GT_{\mu\nu}$), and
the famous work is $f(R)$ theory [4]; On the other hand, we can
modify the material item at the right-handed side of this equation.
Therefore, many workers have always been making an effort to look
for the matter that can lead to this evolutive history of the
universe. By combining cosmic observations, it's implied that an
exotic energy component, which is called dark energy (DE) and has
negative pressure, exists in the universe and its density accounts
for two-thirds of the total energy density. In addition, there have
been some studies which are accomplished to reconcile the two
methods [5]. In the case of the latter, numerous models about DE are
proposed in order to clarify the essence of DE. The simplest one is
the cosmological constant ${\Lambda}$, with the equation of state
(EOS) $w=\frac{P}{\rho}=-1$. However, it confronts with two
difficulties: the fine-tuning problem and the cosmic coincidence
problem. An alternative offer is the dynamical DE to overcome the
difficulties above. In this class of models, the energy form is
provided by an evolving, scalar field ${\phi}$, with a
self-interaction potential $V({\phi})$, such as quintessence [6],
phantom [7], quintom [8] and so forth. Furthermore, there are
generalized Chaplygin gas (GCG) model [9], Holographic DE [10], high
dimension theories [11], etc. Considering more and more DE models,
obviously it is very meaningful to explore the properties of DE in a
model independent manner [12]. Usually, we are able to parameterize
the state parameter $w(z)$ [13], the energy density ${{\rho}(z)}$
[14], or the deceleration parameter $q(z)$ [15] by using
cosmological observations.

In view of the dynamical DE models, we can reconstruct the potential
$V({\phi})$ from supernova observations [16]. The disadvantage of
this kind of methods is that we need to give a concrete form of a
potential, which is usually selected to describe the desired
properties of DE as well as possible, or an expanding form of a
potential. Then, there have been authors reconstructing the
potential from the parameterized parameters without assuming its
form, such as the state parameter $w(z)$ [17], and the energy
density ${{\rho}(z)}$ [18], both of which are used to describe the
properties of DE. When we give parameterized parameters $w(z)$ and
${{\rho}(z)}$, to some extent we have limited some properties of DE.
On the basis of the previous researches, in turn we can think of
reconstructing the potential by a parameter which can directly
reflect the transition from the decelerating expansion to the
accelerating expansion along with the evolving universe. Based on
this consideration, in this paper we will reconstruct the potential
of DE according to the parameterized parameter $q(z)$, being
positive in the past and negative at present with the history of the
universe, which accords with our idea.

Performing a comparison to the previous means, ours has its own
advantages. Firstly, according to the expressions of
$H(z)=H_0\exp\int_0^z(1+q(u))d\ln(1+u)$ and
$H^2(z)=H^2_0[\Omega_{m0}(1+z)^3+(1-\Omega_{m0})(1+z)^{3[1+w(z)]}]$,
we can see that the constructed Hubble parameter $H(z)$ is
independent on cosmological quantity ${\Omega_{m0}}$ from the
deceleration parameter $q(z)$. And it is well known that Hubble
parameter $H(z)$ is an observational quantity, then the cosmological
quantities can be affected by ${\Omega_{m0}}$ when we constrain them
by combining the observational data [19]. But this problem
disappears in the case of the deceleration parameter $q(z)$.
Secondly, it's known that the state parameter $w(z)$ and the energy
density ${{\rho}(z)}$ are made use of describing the properties of
DE. In other words, giving a parameterized state parameter $w(z)$ or
a parameterized energy density ${{\rho}(z)}$ designates that we have
assumed some properties of DE. For our method, we reconstruct the
potential $V({\phi})$ by the parameterized deceleration parameter
$q(z)$ without hypothesizing any evolutive mode of DE and probe DE
completely grounding on the whole cosmological evolvement.
\section{Construction of the potential and the evolution of cosmological parameters}
Considering a spatially flat FRW cosmological model, which consists
of two components: the non-relativistic matter and the dynamical DE,
we utilize a spatially homogeneous scalar field ${\phi}$ to describe
DE.

  With the metric
  $ds^2=-dt^2+a^2(t)[dr^2+r^2(d{\theta}^2+\sin^2{\theta}d{\phi}^2)]$, the Einstein field equation can be written as

\begin{eqnarray}
&&H^2={(\frac{\dot{a}}{a})^2}=\frac{1}{3M_{pl}^2}({\rho}_m+{\rho}_{\phi}),\\
&&\frac{\ddot{a}}{a}=-\frac{1}{6M_{pl}^2}({\rho}_m+{\rho}_{\phi}+3P_{\phi}),
\end{eqnarray}
where, $H=\frac{\dot{a}}{a}$ is the Hubble parameter, ${\rho}_m$ is
the matter density and $M_{pl}\equiv (8{\pi}G)^{-\frac{1}{2}}$ is
the reduced Planck mass. The energy density $\rho_{\phi}$ and the
pressure $P_{\phi}$ of the evolving scalar field $\phi$ are
respectively given by

\begin{eqnarray}
&&\rho_{\phi}=\frac{1}{2}{\dot{\phi}^2}+V({\phi}),\\
&&P_{\phi}=\frac{1}{2}{\dot{\phi}^2}-V({\phi}),
\end{eqnarray}
where, the dot above $\phi$ designates its derivative with respect
to the time t and $V({\phi})$ is the potential of the evolving
scalar field ${\phi}$. From the equations above, we can easily
obtain

\begin{eqnarray}
&&V({\phi})=\frac{1}{2}(\rho_{\phi}-P_{\phi}),\\
&&{\dot{\phi}^2}=\rho_{\phi}+P_{\phi}.
\end{eqnarray}

 Obviously equation (1) can yield
 $\rho_{\phi}=3M_{pl}^2H^2-\rho_m$, while the equation (1) and (2)
 can be combined to result in
 $P_{\phi}=(q-\frac{1}{2})2M_{pl}^2H^2$, where, $q\equiv-\frac{\ddot{a}}{aH^2}$ is the deceleration
 parameter.
  Using $\rho_m=\rho_{m0}(1+z)^3$, where, z is the redshift, and the
relation between the scale factor and the redshift z,
$a(t)=\frac{1}{1+z}$ with $a(t_0)=1$, the equations (5) and (6) can
be rewritten as

\begin{eqnarray}
&&V(\phi(z))=\rho_{m0}[\frac{(2-q)H^2}{3\Omega_{m0}H_0^2}-\frac{1}{2}(1+z)^3],\\
&&(\frac{d\phi}{dz})^2=M_{pl}^2[\frac{2(1+q)}{(1+z)^2}-\frac{3\Omega_{m0}H_0^2}{H^2}(1+z)],
\end{eqnarray}
where, $\Omega_{m0}\equiv \frac{\rho_{m0}}{3M_{pl}^2H_0^2}$ is the
dimensionless energy density, $H_0$ is the Hubble constant and the
subscript 0 denotes the present value of a quantity at the redshift
z=0.

  In order to realize the reconstruction $V({\phi})$ from the
parameterized deceleration parameter $q(z)$, next we must deal with
$H(z)$. The derivative of the Hubble parameter $H(z)$ with regard to
the time t is $\dot{H}=-(1+q)H^2$. Then there exists a contact
between $H(z)$ and $q(z)$ in virtue of an integration:
\begin{eqnarray}
H(z)&&{=}H_0\exp\int_0^z(1+q(u))d\ln(1+u)\nonumber\\
&&{=}H_0f(z)
\end{eqnarray}
where, $f(z)\equiv\exp\int_0^z(1+q(u))d\ln(1+u)$.
  Hence, we are capable of writing down the reconstructed equations for
$V(z)$ and $(\frac{d\phi}{dz})^2$ in term of $q(z)$:
\begin{eqnarray}
&&V(\phi(z))=\rho_{m0}[\frac{(2-q)}{3\Omega_{m0}}f^2(z)-\frac{1}{2}(1+z)^3],\\
&&(\frac{d\phi}{dz})^2=M_{pl}^2[\frac{2(1+q)}{(1+z)^2}-3\Omega_{m0}(1+z)f^{-2}(z)].
\end{eqnarray}

We define the dimensionless quantities
$\tilde{V}\equiv\frac{V}{\rho_{m0}}$ and
$\tilde{\phi}\equiv\frac{\phi}{M_{pl}} $. The reconstructed
equations (10) and (11) can then be ulteriorly written as
\begin{eqnarray}
&&\tilde{V}(\tilde{\phi}(z))=\frac{(2-q)}{3\Omega_{m0}}f^2(z)-\frac{1}{2}(1+z)^3,\\
&&\frac{d\tilde{\phi}}{dz}=-\sqrt{\frac{2(1+q)}{(1+z)^2}-3\Omega_{m0}(1+z)f^{-2}(z)}.
\end{eqnarray}
  These are the main results in our paper, which reveal the
relation between the potential of the scalar field $V({\phi})$ and
the deceleration parameter $q(z)$. If we provide an effective
parameterized deceleration parameter $q(z)$, we can obtain the
reconstruction of the potential $V({\phi})$ by the equations (12)
and (13). Here, we have completed the extraction and chosen the
expression with the minus, namely $\dot{\phi}<0$, in fact which has
no effect to the result. If $\dot{\phi}>0$, we only need to change
the definition of ${\phi}$ to $-{\phi}$.

  In addition, we present the state parameter $w_{\phi}$, the
dimensionless energy density $\Omega_{\phi}$, $\Omega_m$ on $q(z)$:
\begin{eqnarray}
&&w_{\phi}=\frac{P_{\phi}}{\rho_{\phi}}
               =\frac{2q-1}{3[1-\Omega_{m0}(1+z)^3f^{-2}(z)]},\\
&&\Omega_m=\Omega_{m0}(1+z)^3f^{-2}(z),\\
&&\Omega_{\phi}=1-\Omega_m.
\end{eqnarray}

The concrete expressions of these parameters can be readily attained
after we provide an effective parameterized deceleration parameter
$q(z)$. As concerning the parameters in parameterized deceleration
parameter $q(z)$, we are able to constrict them through the latest
observations[20],[21].

 In this paper, we consider two kinds of parameterized
parameters as follows:

  Parametrization 1: $q(z)=a+\frac{bz}{(1+z)^2}$

  Just in order to write conveniently formulae, we define
  $g(z)\equiv(1+z)^{-2(1+a+b)}\exp{[\frac{2bz}{1+z}]}\equiv f^{-2}(z)$. So we have for Parametrization
  1:
\begin{eqnarray}
&&\tilde{V}(z)=\frac{2-a-\frac{b z}{(1+z)^2}}{3\Omega_{m0}}g^{-1}(z)
-\frac{1}{2}(1+z)^3 ,\\
&&\frac{d\tilde{\phi}}{dz}=-\frac{\sqrt{2[1+a+\frac{b z}{(1+z)^2}]-3\Omega_{m0}(1+z)^3g(z)}}{1+z},\\
&&w_{\phi}=\frac{2[a+\frac{b z}{(1+z)^2}]-1}{3[1-\Omega_{m0}(1+z)^3g(z)]},\\
&&\Omega_m=\Omega_{m0}(1+z)^3g(z),\\
&&\Omega_{\phi}=1-\Omega_m.
\end{eqnarray}

 Parametrization 2: $q(z)=\frac{1}{2}+\frac{az+b}{(1+z)^2}$

 It's the same to the case above and we define
 $h(z)\equiv\exp{[-b-\frac{az^2-b}{(1+z)^2}]}$ for convenience, then obtaining
\begin{eqnarray}
&&\tilde{V}(z)=[\frac{\frac{3}{2}-\frac{az+b}{(1+z)^2}}{3\Omega_{m0}h(z)}-\frac{1}{2}](1+z)^3 ,\\
&&\frac{d\tilde{\phi}}{dz}=-\frac{\sqrt{2[\frac{3}{2}+\frac{az+b}{(1+z)^2}]-3\Omega_{m0}h(z)}}{1+z},\\
&&w_{\phi}=\frac{2[\frac{1}{2}+\frac{az+b}{(1+z)^2}]-1}{3[1-\Omega_{m0}h(z)]},\\
&&\Omega_m=\Omega_{m0}h(z),\\
&&\Omega_{\phi}=1-\Omega_m.
\end{eqnarray}

\begin{center}
\includegraphics[width=5.5in, height=2.5in]{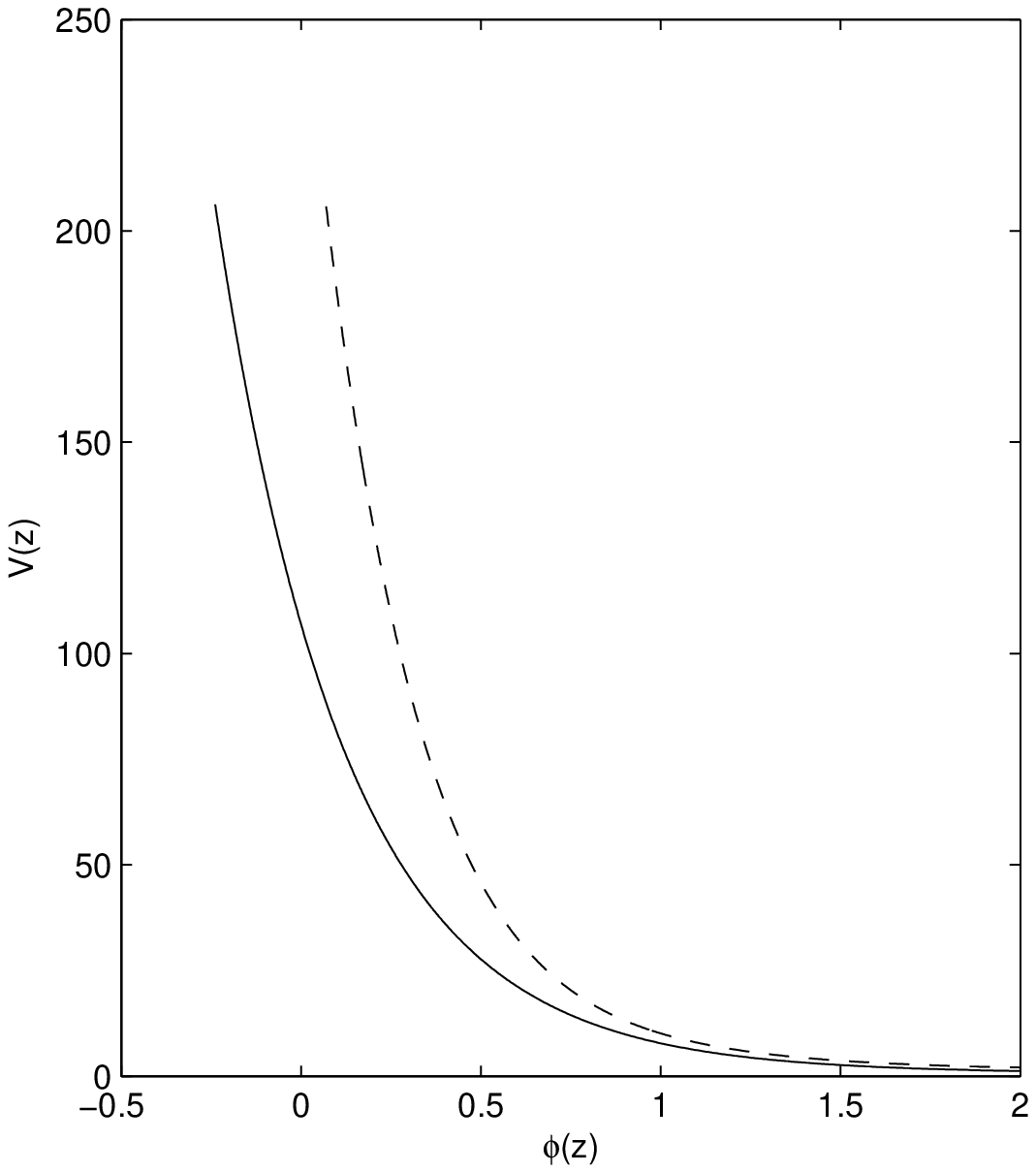}
\textbf{Fig.1} The evolution of the reconstructed potentials with
the scalar field ${\phi}$. \label{figure 1}
\end{center}

 We plot the potentials $V({\phi})$ in
Fig. 1 and the evolution of the cosmological parameters in Fig. 2,
3. We get the values of the parameters by referring to [20] and
taking $\Omega_{m0}=0.3$ as a priority.

\begin{center}
\includegraphics[width=6.5in, height=3in]{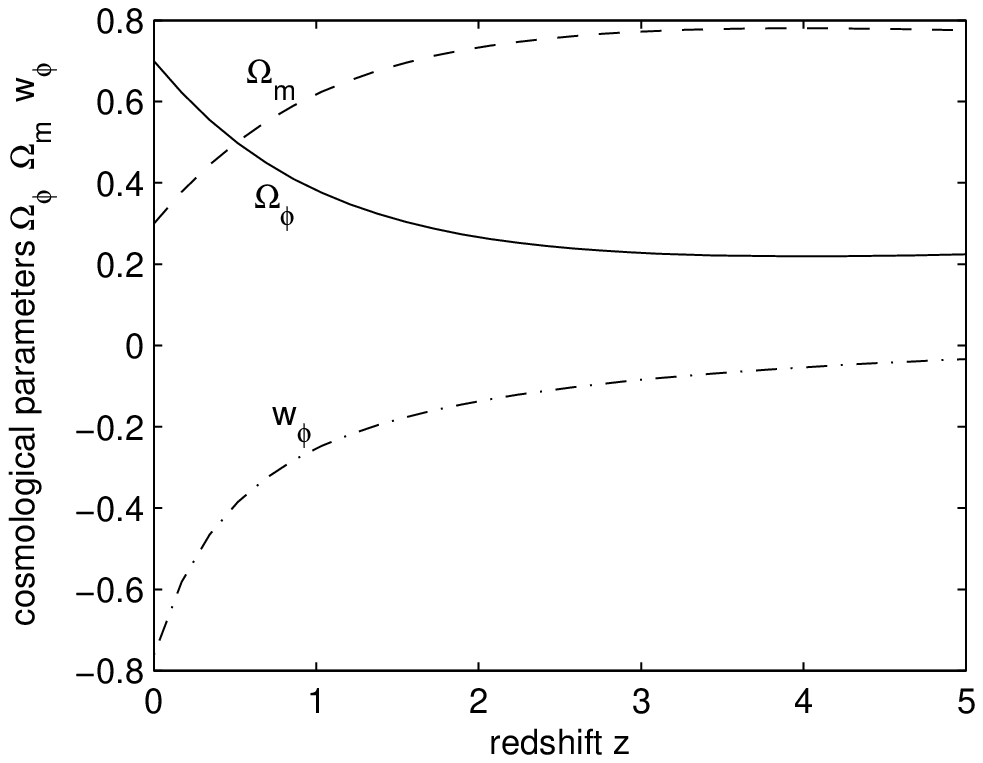}
\textbf{Fig.2} The evolution of the cosmological parameters for
Parametrization 1 with the redshift $z$, where, $a=-0.3$ and $b=0.8$
are used according to the observations constraint. \label{figure 2}
\end{center}

\begin{center}
\includegraphics[width=6.5in, height=3in]{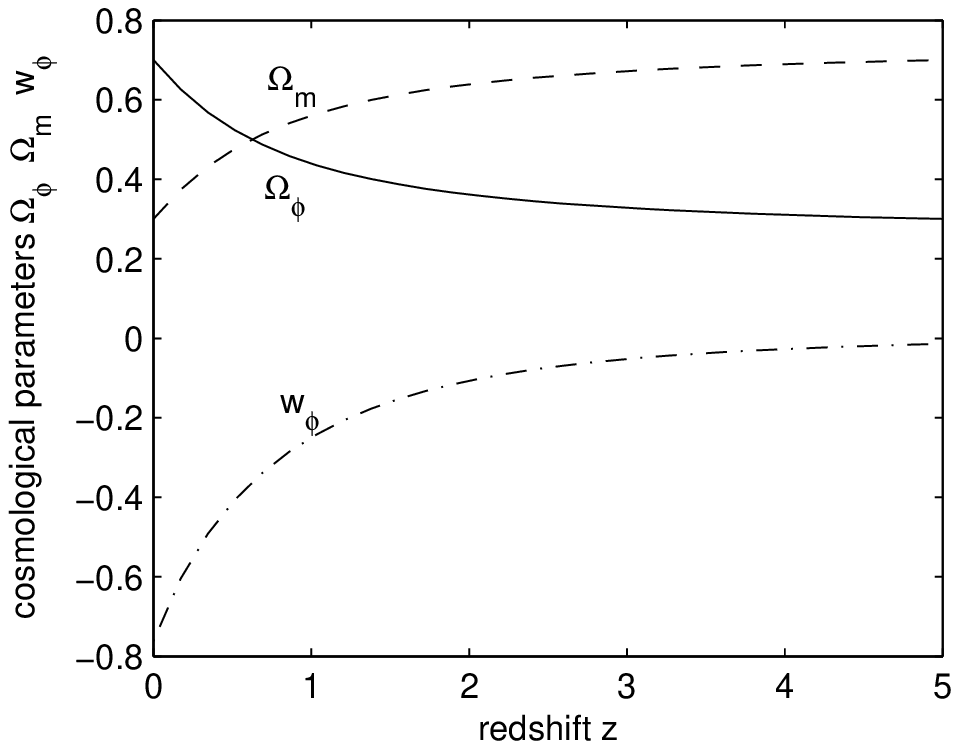}
\textbf{Fig.3} The evolution of the cosmological parameters for
Parametrization 2 with the redshift $z$, where, $a=0.1$ and $b=-0.8$
are used according to the observations constraint. \label{figure 3}
\end{center}

 In theory, we can get the function $\phi(z)$ by integrating the
differential equation (13) for a concrete $q(z)$. Then the
expression of $z$ with respect to $\phi$ can be obtained. After
substituting $z({\phi})$ into the equation (12), we are able to get
the analytical formula of the reconstructed potential. However, as
we have seen above, the differential equations are too intricate to
be integrated analytically in our considered cases. Therefore, at
last only can we numerically evaluate them with the help of the
initial condition $\phi_0=2$. In fact, the value of $\phi_0$ can't
change the shape or trend of the potential $V({\phi})$. It just
leads to shift the graphics horizontally.
 From the Fig. 1, it's demonstrated that the
potentials decrease with the evolving scalar field ${\phi}$ and have
the similar behaviors for the two kinds of examples at the late
period. By our numerical evaluation, we discover that the evolving
scalar field ${\phi}$ linearly runs away as the redshift $z$
increases. So the Figure 1 also presents the change rates of the
reconstructed potential $V({\phi})$ are different between high
redshift and low redshift, the former being faster than the latter.

 Next, we plan to consider a simply form $q=constant$, which is
tenable within a small extension. Inserting it into the equations
(12) and (13), we can hold on by considering discretely at low
redshift and high one. For the low redshift, we can obtain an
approximate analytic exponential expression:
\begin{eqnarray}
&&V(\phi)=\frac{2-q_c}{3\Omega_{m0}}\exp{[\frac{-(2q_c-1)(\phi-\phi_0)}{\sqrt{2(1+q_c)-3\Omega_{m0}}}]}
\end{eqnarray}
At high redshift, the result is different from that at low redshift.
The formula is in square law:
\begin{eqnarray}
&&V(\phi)=\frac{(2-q_c)(q_c-\frac{1}{2})^2}{4(1+q_c)^4}(\phi-\phi_0)^2
\end{eqnarray}
It's found that this expression is independent on the value of the
dimensionless matter density $\Omega_{m0}$.

\section{Conclusions}
In summary, the dark energy potentials can be reconstructed from
parameterized deceleration parameters. For the two examples given
above, we obtain the evolution of the reconstructed potential
$V(\phi)$, the dimensionless energy density $\Omega_{\phi}$,
$\Omega_{m}$ and the state parameter $w_\phi$ for a given
parameterized deceleration parameter $q(z)$. The result shows that
the reconstructed potentials $V(\phi)$ decrease with the evolving
scalar field $\phi$, namely that the reconstructed potentials
$V(\phi)$ have the direction of going down with the expansion of the
universe. Also it's found that the change rates of the reconstructed
potential $V(\phi)$ are obviously diverse, which at $\phi<1$ is
quicker than that at $\phi>1$. For the two cases, the variational
trends of the reconstructed potentials $V(\phi)$ are very similar at
the late period. In addition, analytically evaluating the equations
(12) and (13) with $q=constant$, we obtain two different approximate
expressions and find the formula (28) don't contain the
dimensionless matter density $\Omega_{m0}$. As far as the theoretic
view is concerned, we establish a relation between the deceleration
parameter and the effective field theory in our means. With the
progress of the observation techniques in the future, the parameters
in $q(z)$ can be constrained more precisely, which can go far
towards understanding the nature of dark energy for us. \emu \bc
\rule{8cm}{0.1mm} \ec \bmu \vs{3.5mm}

\emu\label{last-page}
\end{document}